\documentstyle[epsfig]{article}
\begin{document}
\begin{center}
{\LARGE{\bf A $q$-deformed nonlinear map }} \\  

\vspace{1cm}

Ramaswamy Jaganathan\footnote{jagan@imsc.res.in} and 
Sudeshna Sinha\footnote{sudeshna@imsc.res.in}\\ 
{\it The Institute of Mathematical Sciences,\\ 
Tharamani, Chennai 600 113, India}\\
\end{center}

\bigskip

\begin{abstract}
A scheme of $q$-deformation of nonlinear maps is introduced.  As a specific 
example, a $q$-deformation procedure related to the Tsallis $q$-exponential 
function is applied to the logistic map. Compared to the canonical logistic 
map, the resulting family of $q$-logistic maps is shown to have a wider 
spectrum of interesting behaviours, including the co-existence of attractors 
-- a phenomenon rare in one dimensional maps.
\end{abstract}

\bigskip


\bigskip

\section{Introduction} 
Emergence of the so-called quantum group structures in certain physical 
problems has led to studies on several $q$-deformed physical 
systems~\cite{QG}.  Enthused over this, and inspired by the elements of 
Tsallis statistical mechanics~\cite{TS}, we suggest a scheme of 
$q$-deformation of nonlinear maps. We then elucidate the general features 
of a $q$-deformed logistic map related to the Tsallis $q$-exponential 
function, as a concrete illustration of the scheme of $q$-deformation of 
nonlinear maps.

Theory of quantum groups turned the attention of physicists to the rich 
mathematics of $q$-series, $q$-special functions, etc., with a history 
going back to the nineteenth century~\cite{GR}. The $q$-deformation of 
any function involves essentially a modification of it such that in the 
limit $q \longrightarrow 1$ the original function is recovered. Thus there 
exist several $q$-deformations of the same function introduced in different 
mathematical and physical contexts.  Here, we are concerned mainly with 
the $q$-deformation of real numbers and the exponential function.

Originally, in 1846 Heine deformed a number to a basic number as 
\begin{equation}
[n]_q = \frac{1-q^n}{1-q},
\label{eq:heinen}
\end{equation} 
such that $[n]_q \longrightarrow n$ when $q \longrightarrow 1$.  In 1904 
Jackson defined a $q$-exponential function given by 
\begin{equation}
E_q(x) = \sum_{n=0}^\infty \frac{x^n}{[n]_q!},
\label{eq:qexp}
\end{equation} 
with 
\begin{equation}
[n]_q! = [n]_q[n-1]_q\cdots[2]_q[1]_q, \qquad [0]_q! = 1, 
\end{equation}
as the the solution of the $q$-differential equation 
\begin{equation}
\frac{df(x)}{d_qx} = \frac{f(x)-f(qx)}{(1-q)x} = f(x). 
\end{equation}
It is seen that $E_q(x) \longrightarrow \exp(x)$ in the limit 
$q \longrightarrow 1$ when the Jackson $q$-differential operator $d/d_qx$ 
also becomes the usual differential operator $d/dx$.   

The mathematics of quantum groups necessitated a new deformation of number as 
\begin{equation}
[n]_q = \frac{q^n-q^{-n}}{q-q^{-1}},
\label{eq:qgn}
\end{equation} 
which also has the required property that in the limit $q \longrightarrow 1$,
$[n]_q \longrightarrow n$.  The associated $q$-exponential function is given 
by the same equation~(\ref{eq:qexp}) but with $[n]_q$ defined according 
to~(\ref{eq:qgn}).

In the nonextensive statistical mechanics of Tsallis~\cite{TS}, a new  
$q$-exponential function has been introduced as given by 
\begin{equation}
e_q^x = (1+(1-q)x)^{1/(1-q)},
\label{eq:tqexp}
\end{equation}
which satisfies the nonlinear equation 
\begin{equation} 
\frac{df(x)}{dx} = (f(x))^q,  
\end{equation} 
and has the required limiting behaviour\,: $e_q^x \longrightarrow \exp(x)$  
when $q \longrightarrow 1$.  This $e_q^x$ plays a central role in the 
nonextensive statistical mechanics by replacing $\exp(x)$ in certain domains 
of application; it should be noted that it is natural to define a generalized 
exponential function as in~(\ref{eq:tqexp}) if we consider the relation 
\begin{equation}
e^x = \lim_{N\longrightarrow\infty}\left(1+\frac{x}{N}\right)^N, 
\end{equation} 
and regard $1/N$ as a continuous parameter.  The formalism of nonextensive 
statistical mechanics has found applications in a wide range of physical 
problems~\cite{TS}, including the study of nonlinear maps at the edge of 
chaos. Here we derive another deformation of numbers based on the Tsallis 
$q$-exponential function defined by~(\ref{eq:tqexp}), and use it to study 
a $q$-deformed logistic map as an example of the general scheme of 
$q$-deformation of nonlinear maps.

\section{A $q$-deformation scheme for nonlinear maps} 
The series expansion of $e_q^x$ has been presented in~\cite{B} as 
\begin{equation}
e_q^x = 1+\sum_{n=1}^\infty \frac{Q_{n-1}x^n}{n!},
\label{eq:bseries}
\end{equation} 
with 
\begin{equation}
Q_n = 1(q)(2q-1)(3q-2)\cdots(nq-(n-1)), \qquad n = 0,1,2,\cdots\,.
\end{equation}
Let us take 
\begin{equation}
1-q = \epsilon,
\end{equation} 
and write 
\begin{equation}
e_q^x = \tau_\epsilon(x) = (1+\epsilon x)^{1/\epsilon}.  
\end{equation}
From~(\ref{eq:bseries}) the series expansion of $\tau_\epsilon(x)$
follows as 
\begin{equation}
\tau_\epsilon(x) = \sum_{n=0}^\infty \frac{T_n x^n}{n!},
\label{eq:tauseries}
\end{equation} 
with
\begin{eqnarray}
T_n & = & \left\{
          \begin{array}{ll}
          1, & \qquad {\mbox{for}}\ n = 0, \\
          1(1-\epsilon)(1-2\epsilon)\cdots(1+(1-n)\epsilon), 
             & \qquad {\mbox{for}}\ n\geq 1.
          \end{array} \right. \nonumber \\
  &    &   
\end{eqnarray}
Comparison of~(\ref{eq:qexp}) and~(\ref{eq:tauseries}) suggests that
we have here another deformation of numbers\,:  
\begin{equation}
[n]_\epsilon = \frac{n}{1+(1-n)\epsilon},
\label{eq:defn}
\end{equation} 
such that
\begin{equation}
\lim_{\epsilon \longrightarrow 0}[n]_\epsilon = n.
\end{equation} 
Then, we have 
\begin{equation}
\tau_\epsilon(x) = \sum_{n=0}^\infty \frac{x^n}{[n]_\epsilon !},
\end{equation} 
exactly analogous to the expression for $E_q(x)$ in~(\ref{eq:qexp}). 
 
It is usual to extend the deformation rule for integers, such as 
in~(\ref{eq:heinen}) or~(\ref{eq:qgn}), to any $X$ by substituting $X$ 
for $n$.  For example, the commutation relations of the $q$-deformed $su(2)$ 
algebra are given by 
\begin{equation} 
J_0J_\pm-J_\pm J_0 = \pm J_0, \qquad
J_+J_--J_-J_+ = [2J_0]_q,
\end{equation}
where $[2J_0]_q$ is obtained from the definition of $[n]_q$ in~(\ref{eq:qgn}) 
by replacing $n$ by $2J_0$, i.e.,  
$[2J_0]_q = (q^{2J_0}-q^{-2J_0})/(q-q^{-1})$.  Thus, we take that for any 
real number $x$ 
\begin{equation} 
[x]_\epsilon = \frac{x}{1+\epsilon(1-x)}.  
\label{eq:xq}
\end{equation} 
Note that $[x]_\epsilon \longrightarrow x$ when $\epsilon \longrightarrow 0$.  
Further, $[0]_\epsilon = 0$ and $[1]_\epsilon = 1$, as in the case of other 
deformations in~(\ref{eq:heinen}) and~(\ref{eq:qgn}).  Here after, we shall 
denote $[x]_\epsilon$ simply as $[x]$ and take it to be defined 
by~(\ref{eq:xq}) unless stated otherwise.

The $q$-deformation scheme for discrete dynamical maps we suggest is the 
following.  For the one-dimensional map 
\begin{equation}
x_{n+1} = f(x_n) 
\end{equation} 
the $q$-deformed version is 
\begin{equation}
x_{n+1} = f([x_n]_q), 
\end{equation} 
where $[x_n]_q$ is, in general, any $q$-deformed value of $x_n$; essentially 
a $q$-deformed map is obtained by composing the given map with a basic number 
deforming map. For example, the $q$-deformed logistic map, corresponding 
to the definition of deformed $x$ as in~(\ref{eq:xq}), is\,:
\begin{equation}
x_{n+1} = a[x_n](1-[x_n]) = F(x_n) 
        = \frac{a(1+\epsilon)x_n(1-x_n)}{(1+\epsilon(1-x_n))^2}.
\label{eq:qlog}
\end{equation} 
We shall refer to the map in~(\ref{eq:qlog}) simply as the $q$-logistic
map.  When $\epsilon \longrightarrow 0$ the $q$-logistic map becomes the usual 
logistic map.  In the following we present the interesting properties of the 
$q$-logistic map in detail.  

\section{The $q$-logistic map}
Figure~1 displays the $q$-logistic map function in~(\ref{eq:qlog}), 
\begin{equation}
F(x) = \frac{a(1+\epsilon)x(1-x)}{(1+\epsilon(1-x))^2}, 
\label{eq:qmapf} 
\end{equation} 
for $a = 4$.  The usual logistic map corresponds to $\epsilon = 0$.  It is
clear that as $\epsilon$ moves away from $0$ the map gets more skewed.  The
map is skewed to the right for positive $\epsilon$ and to the left for
negative $\epsilon$.  The lower and upper bounds for $\epsilon$ are, 
respectively, $-1$ and $\infty$ where $F(x)$ vanishes.  Taking the domain 
of the $q$-logistic map the same as for the logistic map, namely the interval
$[0,1]$, it is found that the range of the $q$-logistic map is the same as 
for the logistic map.  This is so because over the interval $[0,1]$ the map 
function $F(x)$ has no singularity for $-1<\epsilon<\infty$, and $x \leq [x]\leq 1$ 
for $-1\leq\epsilon<0$ and $[x]\leq x$ for $\epsilon\geq 0$.  There is one 
important qualitative difference between the usual logistic map and its 
deformed version. {\em The deformed map is concave in parts of $x$-space, 
whereas the usual logistic map is always convex}.  Note that the form of the 
$q$-logistic map is similar to certain maps used to model population dynamics, 
such as the Bellows map\,: $f(x) = rx/(1+x^b)$, and the Moran-Ricker 
exponential map\,: $f(x) = x\exp\{r(1-x)\}$, where $r$ is the nonlinearity 
parameter leading from periodic behavior to chaos~\cite{population}.

\begin{figure}[htb] 
\label{fig1}
\begin{center}
  \mbox{\epsfig{file=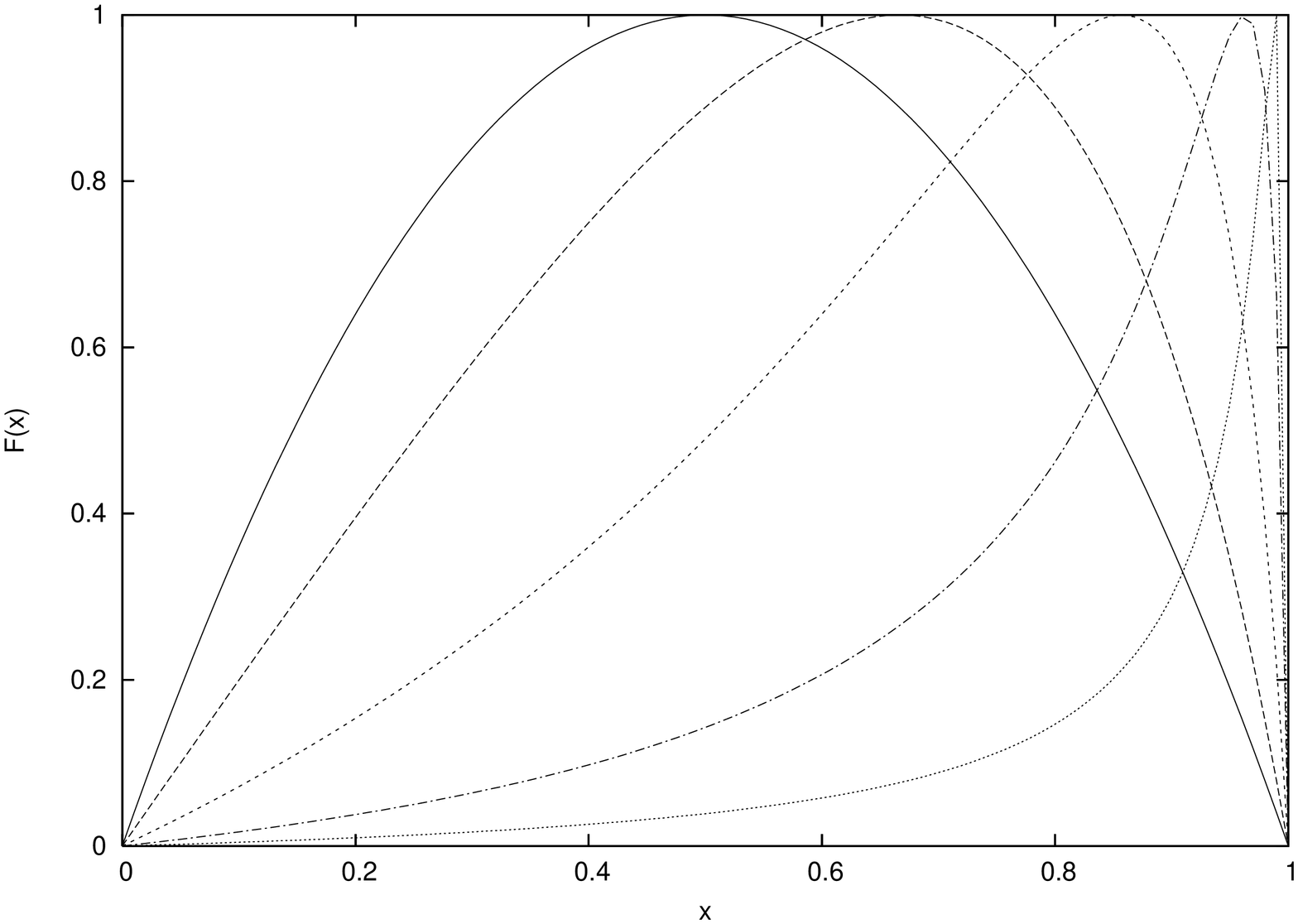,width=5.5truecm,angle=-90}}
  \mbox{\epsfig{file=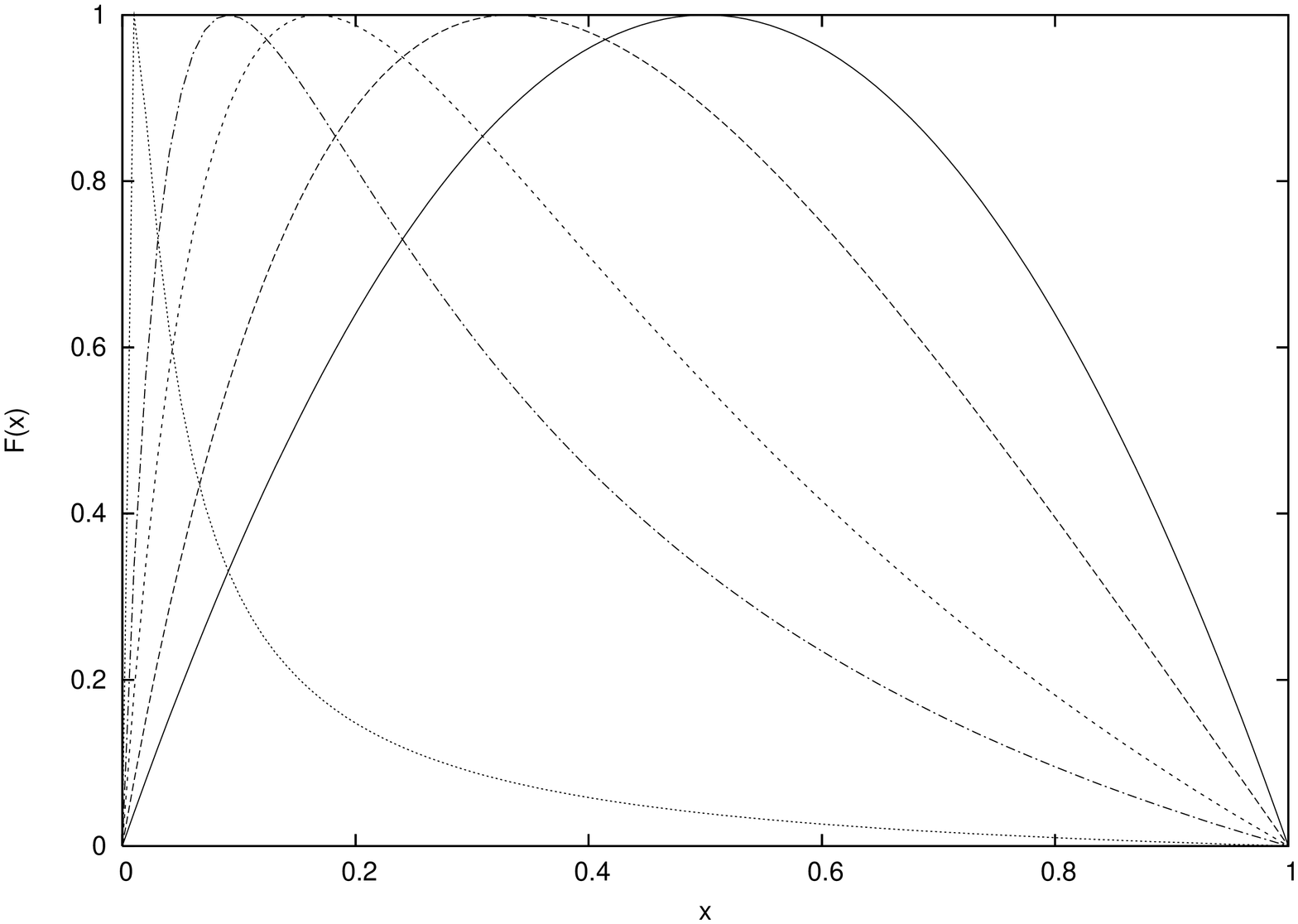,width=5.5truecm,angle=-90}}
\end{center} 
\caption{Graph of the $q$-logistic map function $F(x)$ for $a = 4$. The solid 
  curve is for $\epsilon = 0$ (usual logistic map). The dashed curves skewed 
  increasingly to the right in the top panel are for $\epsilon = 1,5,25,100$.  
  The dashed curves skewed increasingly to the left in the bottom panel are 
  for $\epsilon = -0.5,-0.8,-0.9,-0.99$.} 
\end{figure} 

The fixed points of the map in~(\ref{eq:qlog}) are given by 
\begin{equation}
x^\star = \frac{a(1+\epsilon)x^\star(1-x^\star)}
               {(1+\epsilon(1-x^\star))^2}. 
\end{equation} 
This has one solution at $x^\star = 0$.  For $x^\star \neq 0$ the above 
equation becomes 
\begin{equation}
1 = \frac{a(1+\epsilon)(1-x^\star)}
         {(1+\epsilon(1-x^\star))^2}. 
\end{equation} 
Substituting $1-x^\star = y$ in this equation gives 
\begin{equation}
(1+\epsilon y)^2 = a(1+\epsilon)y, 
\end{equation} 
or 
\begin{equation}
\epsilon^2y^2+(2\epsilon-a(1+\epsilon))y+1 = 0.  
\end{equation} 
So 
\begin{equation}
y = \frac{1}{2\epsilon^2}
    \left\{a(1+\epsilon)-2\epsilon\pm
    \sqrt{a^2(1+\epsilon)^2-4a\epsilon(1+\epsilon)}\right\}   
\end{equation} 
Thus, the fixed points $\{x^\star\}$ of the $q$-logistic map are\,: 
\begin{eqnarray}
x^\star & = & 0, \\
x^\star_+ & = & \left(1-\frac{1}{2\epsilon^2}
                \left\{a(1+\epsilon)-2\epsilon\right\}\right) 
                +\frac{1}{2\epsilon^2}\sqrt{a^2(1+\epsilon)^2
                                       -4a\epsilon(1+\epsilon)}, \\ 
x^\star_- & = & \left(1-\frac{1}{2\epsilon^2}
                \left\{a(1+\epsilon)-2\epsilon\right\}\right)  
                -\frac{1}{2\epsilon^2}\sqrt{a^2(1+\epsilon)^2
                                       -4a\epsilon(1+\epsilon)}.  
\label{eq:xminus}
\end{eqnarray} 
Figure~2 displays the numerically obtained fixed points of the map for 
$a = 1.5$ for different values of $\epsilon$.  These coincide exactly with
the analytical expression for $x^\star_+$ above.  It should be noted that
for a given value of $a$ only for some range of values of $\epsilon$ both 
$x^\star_+$ and $x^\star_-$ lie within the range of $F(x)$.  

\begin{figure}[htb] 
\label{fig2}
\begin{center}
  \mbox{\epsfig{file=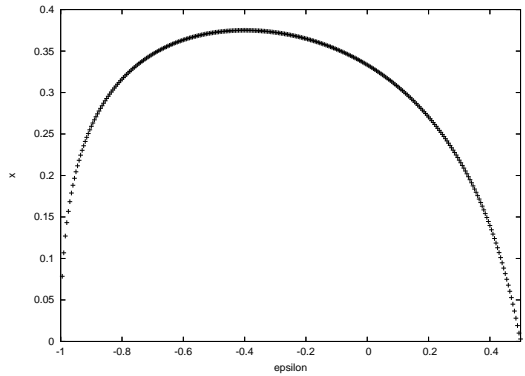,width=5truecm,angle=-90}}
\end{center}
\caption{Bifurcation diagram of the map with respect to $\epsilon$ for 
$a = 1.5$} 
\end{figure} 

Figure~3 gives the bifurcation diagram of the map for $a = 3$ with resepct to 
different values of $\epsilon$.  It is clear that the fixed point $x^\star_+$ 
loses stability for negative $\epsilon$.  This can be straight-forwardly 
understood from the absolute magnitude of $F^\prime(x^\star)$, which crosses 
the value $1$ at $\epsilon = 0$ when $a = 3$.

\begin{figure}[htb]
\label{fig3}
\begin{center}
  \mbox{\epsfig{file=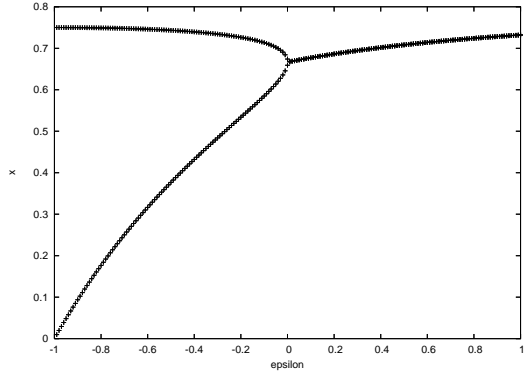,width=5truecm,angle=-90}}
\end{center}
\caption{Bifurcation diagram of the map with respect to $\epsilon$ for 
$a = 3$} 
\end{figure}

Figure~4 gives the bifurcation diagram of the map for $a = 3.5$ with respect 
to different values of $\epsilon$.  In this case the usual logistic map 
($\epsilon = 0$) yields a $4$-cycle.  As $\epsilon$ increases there is a 
reverse bifurcation and the $2$-cycle gains stability, followed by the fixed 
point.  However, note that the fixed point at $x^\star = 0$ also gains 
stability when $\epsilon$ is sufficiently high, and we have a 
{\em co-existence} of attractors, namely, the fixed point at $0$ co-exists, 
first with the $2$-cycle and then with the fixed point $x^\star_+$.

\begin{figure}[htb]
\label{fig4}
\begin{center}
  \mbox{\epsfig{file=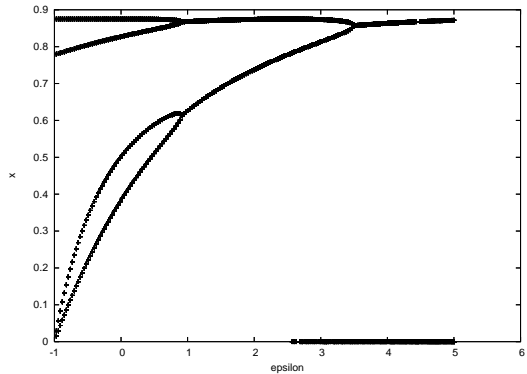,width=5truecm,angle=-90}}
\end{center}
\caption{Bifurcation diagram of the map with respect to $\epsilon$ for 
$a = 3.5$} 
\end{figure}

Similar features, namely reverse bifurcations in $\epsilon$-space, and the 
co-existence of the fixed point $x^\star = 0$ with other dynamical 
behaviour at high $\epsilon$, are observed for larger values of $a$ as
well (see Figs.~5-9). 

\begin{figure}[htb]
\label{fig5}
\begin{center}
  \mbox{\epsfig{file=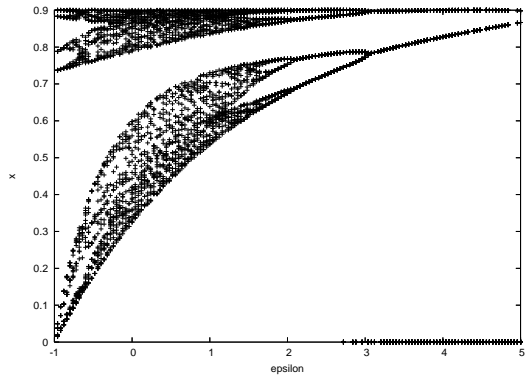,width=5truecm,angle=-90}}
\end{center}
\caption{Bifurcation diagram of the map with respect to $\epsilon$ for 
$a = 3.6$} 
\end{figure}

\begin{figure}[htb]
\label{fig6}
\begin{center}
  \mbox{\epsfig{file=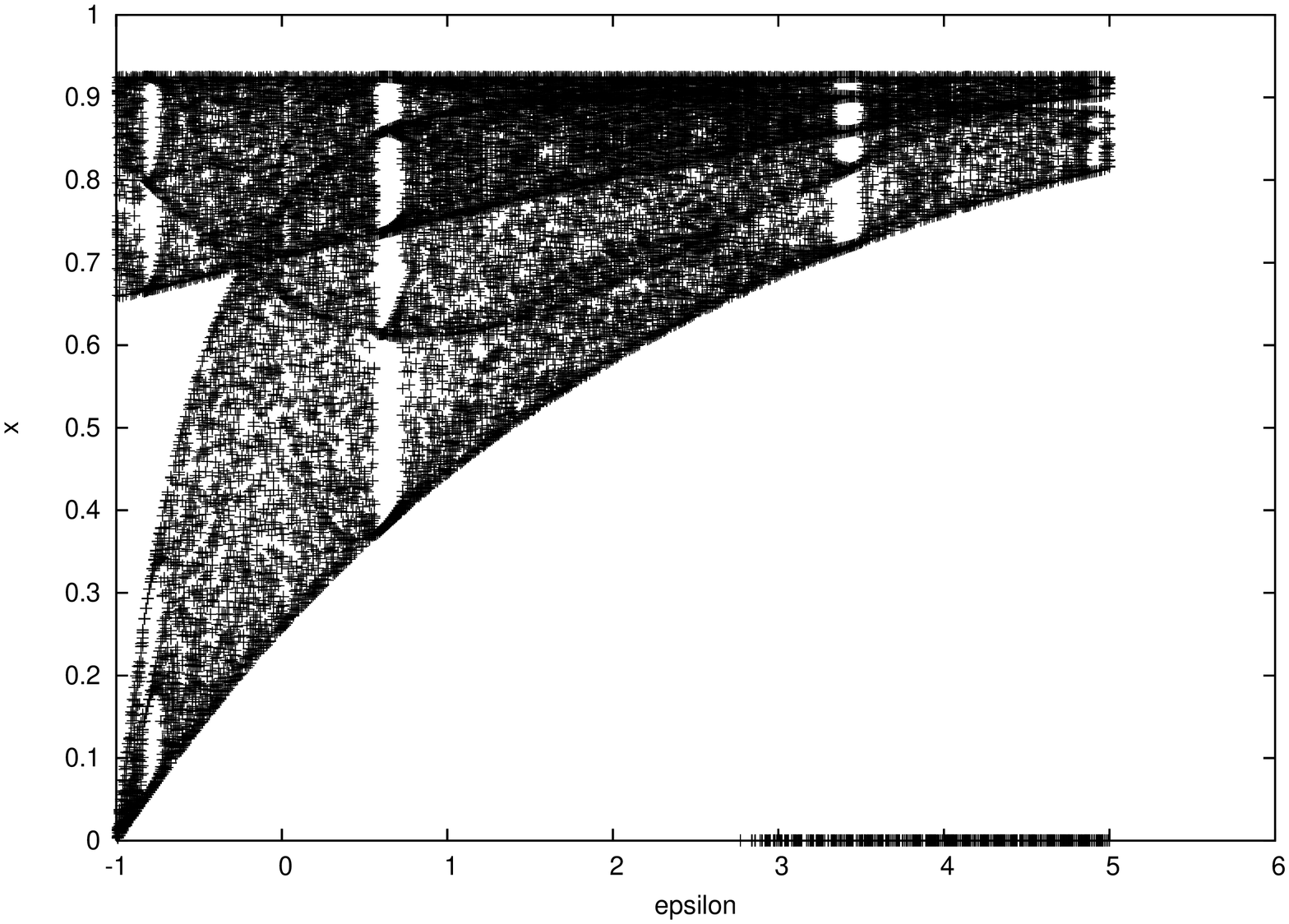,width=5truecm,angle=-90}}
\end{center}
\caption{Bifurcation diagram of the map with respect to $\epsilon$ for 
$a = 3.7$} 
\end{figure}

\begin{figure}[htb]
\label{fig7}
\begin{center}
  \mbox{\epsfig{file=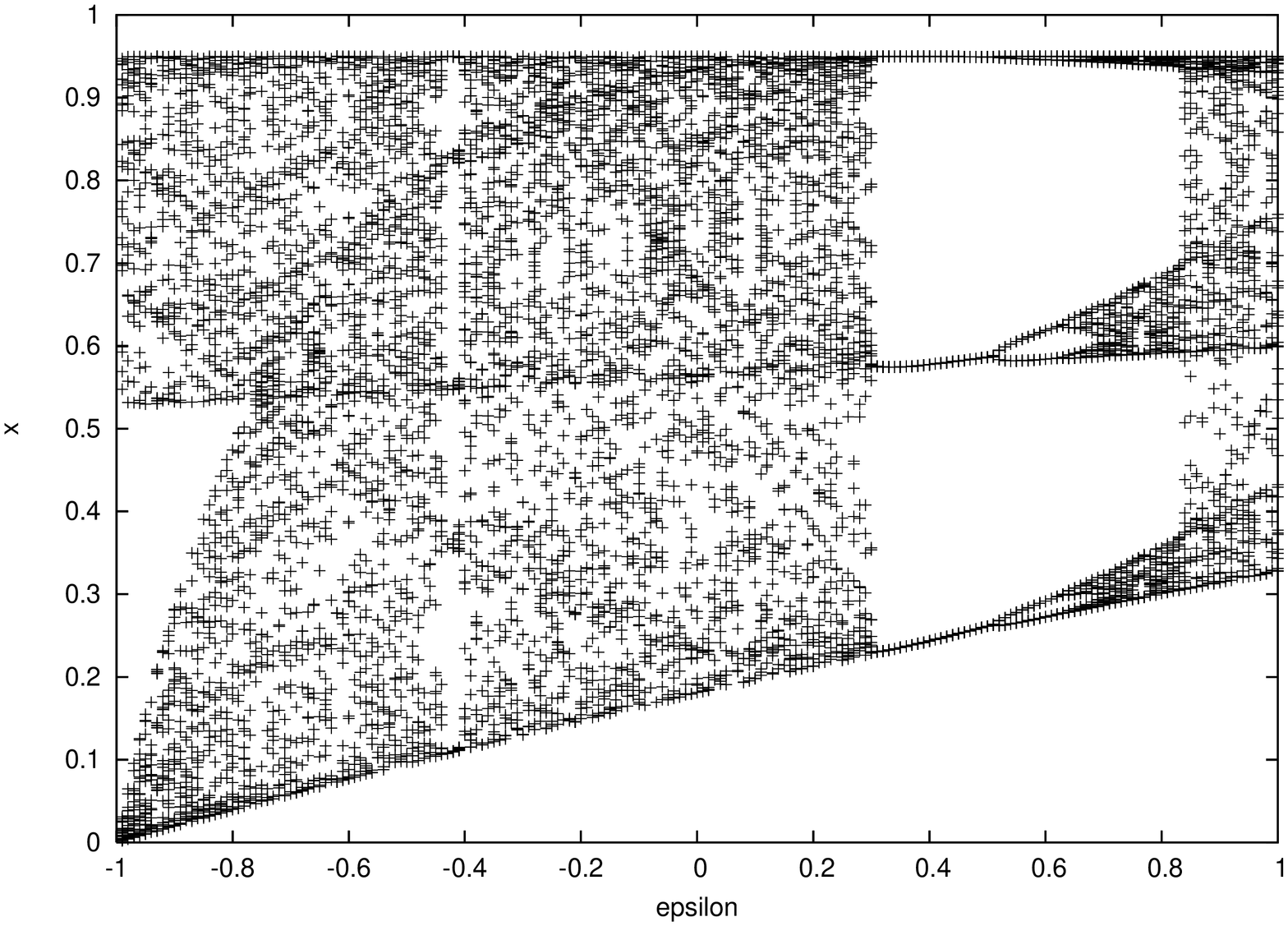,width=5truecm,angle=-90}}
\end{center}
\caption{Bifurcation diagram of the map with respect to $\epsilon$ for 
$a = 3.8$} 
\end{figure}

\begin{figure}[htb]
\label{fig8}
\begin{center}
  \mbox{\epsfig{file=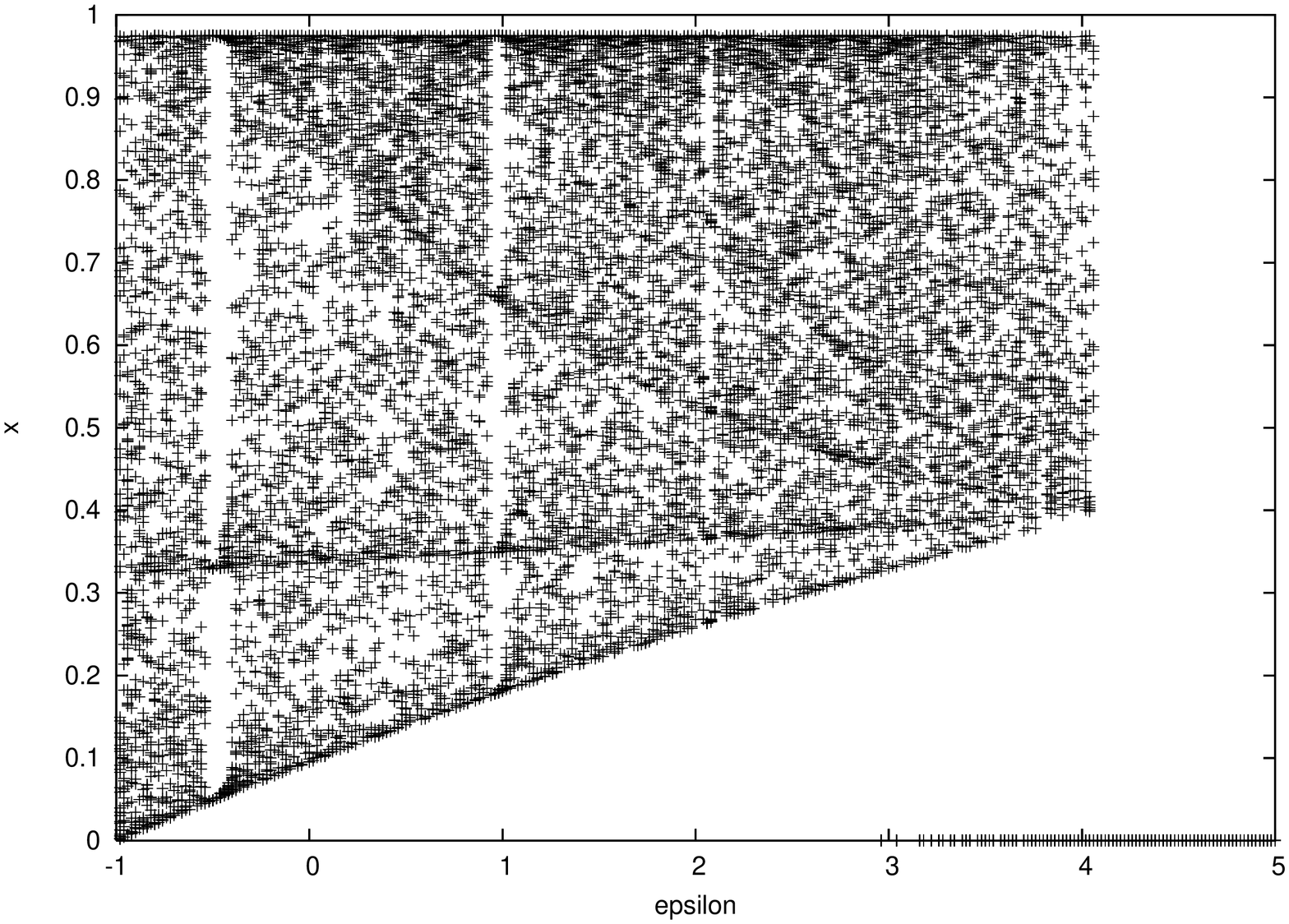,width=5truecm,angle=-90}}
\end{center}
\caption{Bifurcation diagram of the map with respect to $\epsilon$ for 
$a = 3.9$} 
\end{figure}

\begin{figure}[htb]
\label{fig9}
\begin{center}
  \mbox{\epsfig{file=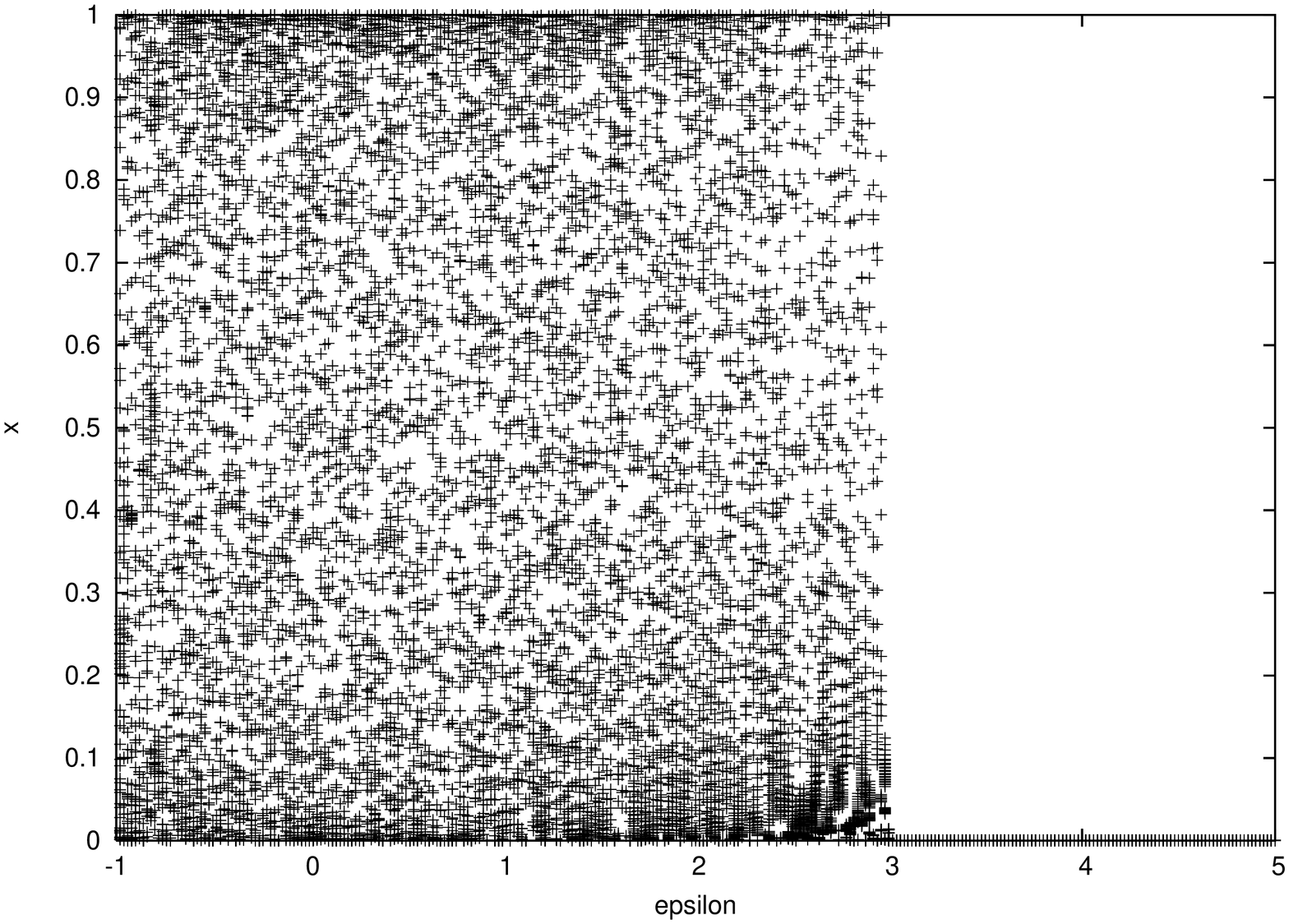,width=5truecm,angle=-90}}
\end{center}
\caption{Bifurcation diagram of the map with respect to $\epsilon$ for 
$a = 4$} 
\end{figure}

\section{Co-existence of attractors}  
It is clearly evident from the above that the $q$-logistic map offers a rare 
example of multiple attractors in a one-dimensional smooth unimodal  
system~\cite{H}.  The fixed point at $x^\star = 0$ co-exists with other kinds 
of dynamical behaviour when $\epsilon$ is sufficiently high.  The basin of 
attraction for $x^\star = 0$, as reflected in the fraction of initial 
conditions which are attracted to $0$, monotonically increases with 
$\epsilon$.  Figures~10-12 show the basins of attraction for the fixed point 
$x^\star = 0$ for three different values of $a$.  It is clear that this fixed 
point is a global attractor at sufficiently high $\epsilon$.  In Figure~12 
for $a = 4$ clearly there is a sharp transition to a global attractor (where 
all initial conditions lead to $x^\star = 0$) at around $\epsilon \sim 3$.  
So, the fully chaotic logistic map under deformation with positive $\epsilon$ 
can yield stable fixed points.

\begin{figure}[htb]
\label{fig10}
\begin{center}
  \mbox{\epsfig{file=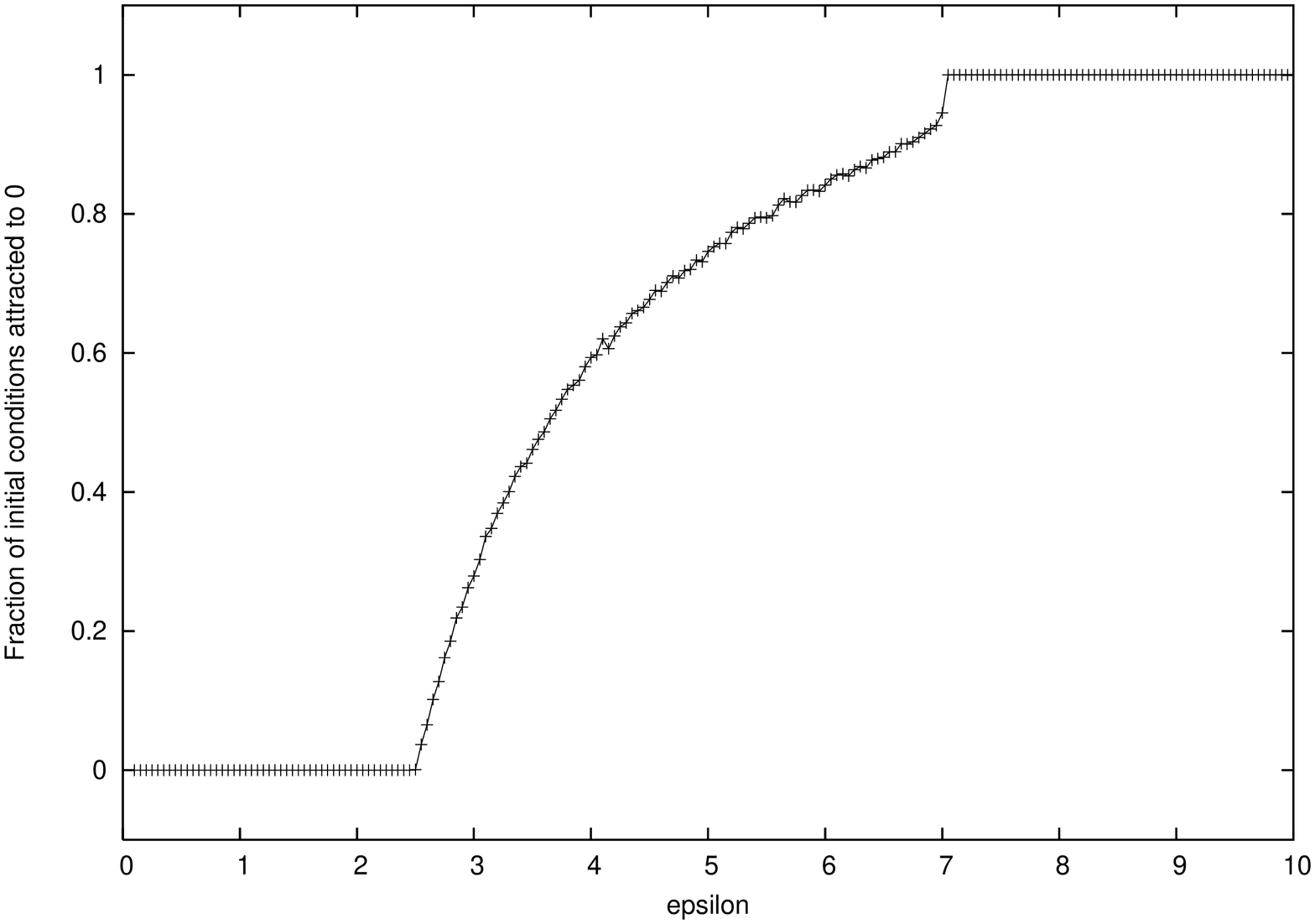,width=5truecm,angle=-90}}
\end{center}
\caption{Fraction of initial conditions attracted to the fixed point 
$x^\star = 0$, for varying values of $\epsilon$ for $a = 3.5$.}
\end{figure}

\begin{figure}[htb]
\label{fig11}
\begin{center}
  \mbox{\epsfig{file=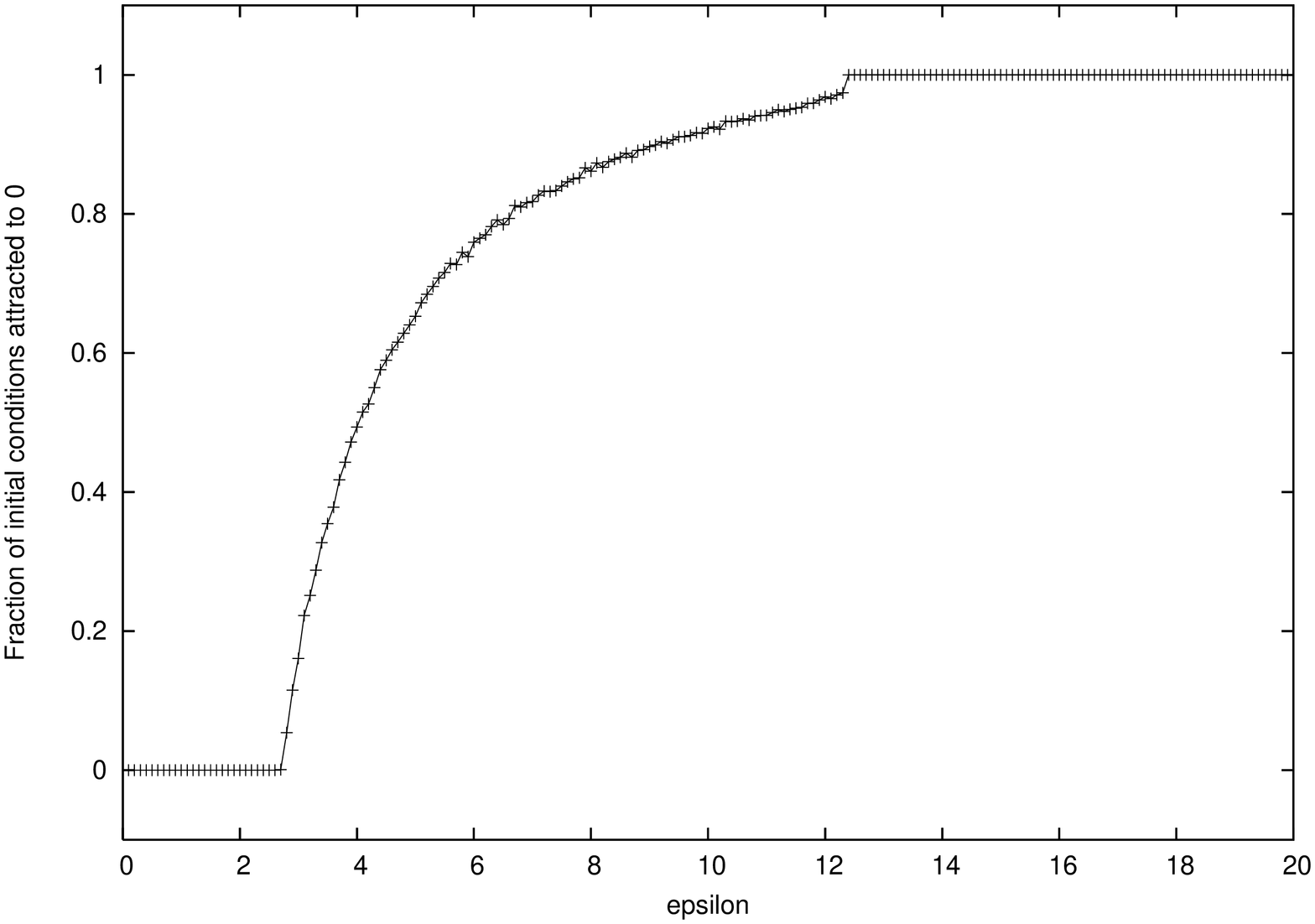,width=5truecm,angle=-90}}
\end{center}
\caption{Fraction of initial conditions attracted to the fixed point 
$x^\star = 0$, for varying values of $\epsilon$ for $a = 3.7$.}
\end{figure}

\begin{figure}[htb]
\label{fig12}
\begin{center}
  \mbox{\epsfig{file=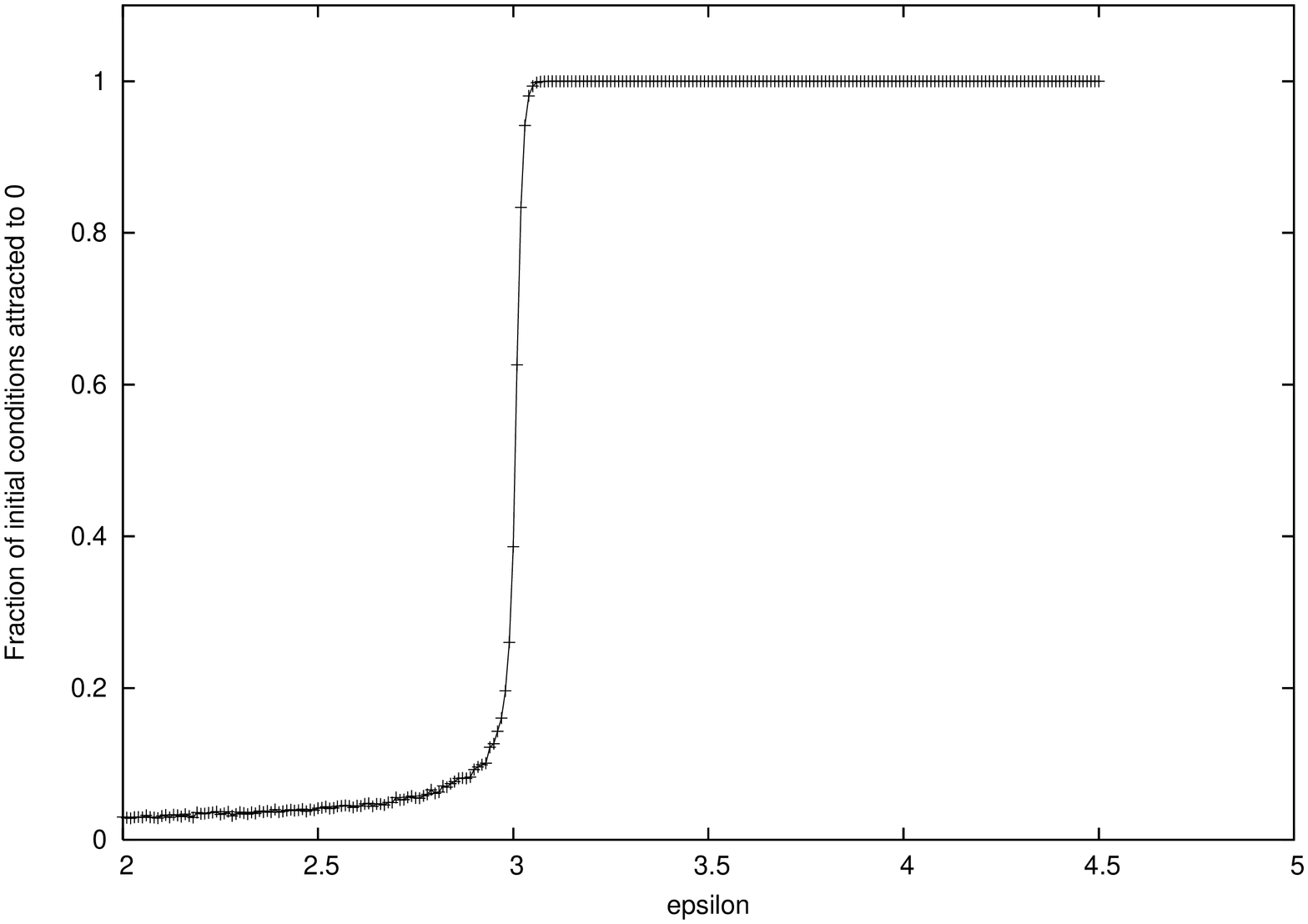,width=5truecm,angle=-90}}
\end{center}
\caption{Fraction of initial conditions attracted to the fixed point 
$x^\star = 0$, for varying values of $\epsilon$ for $a = 4$.} 
\end{figure} 

The co-existence of attractors is also evident from the bifurcation diagrams 
for the map with respect to $a$ for different values of $\epsilon$ displayed 
in Figs.~13-15.  

\begin{figure}[htb]
\label{fig13}
\begin{center}
\mbox{\epsfig{file=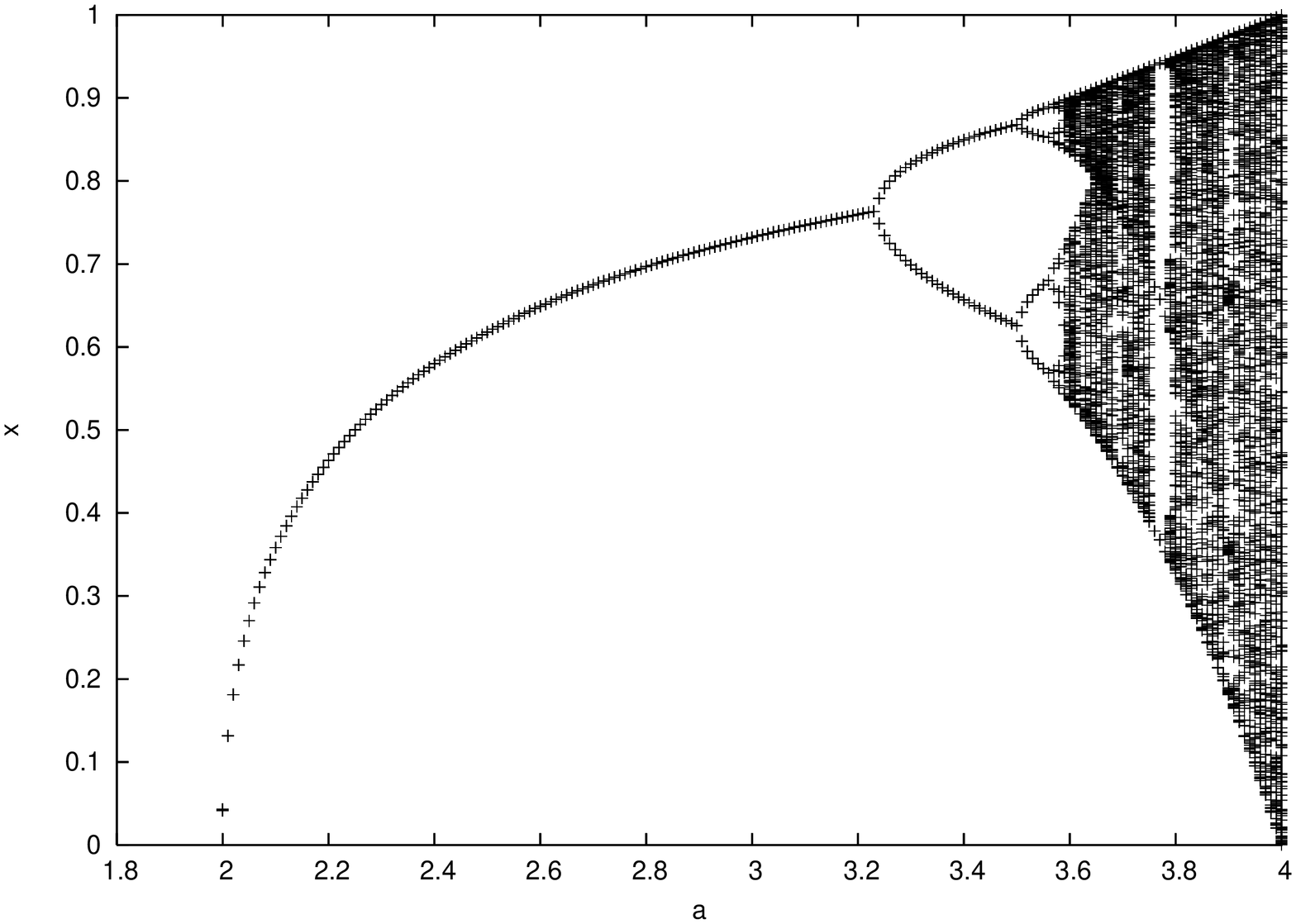,width=5truecm,angle=-90}}
\end{center}
\caption{Bifurcation diagram of the map with respect to $a$ for $\epsilon = 1$}
\end{figure}

\begin{figure}[htb]
\label{fig14}
\begin{center}
  \mbox{\epsfig{file=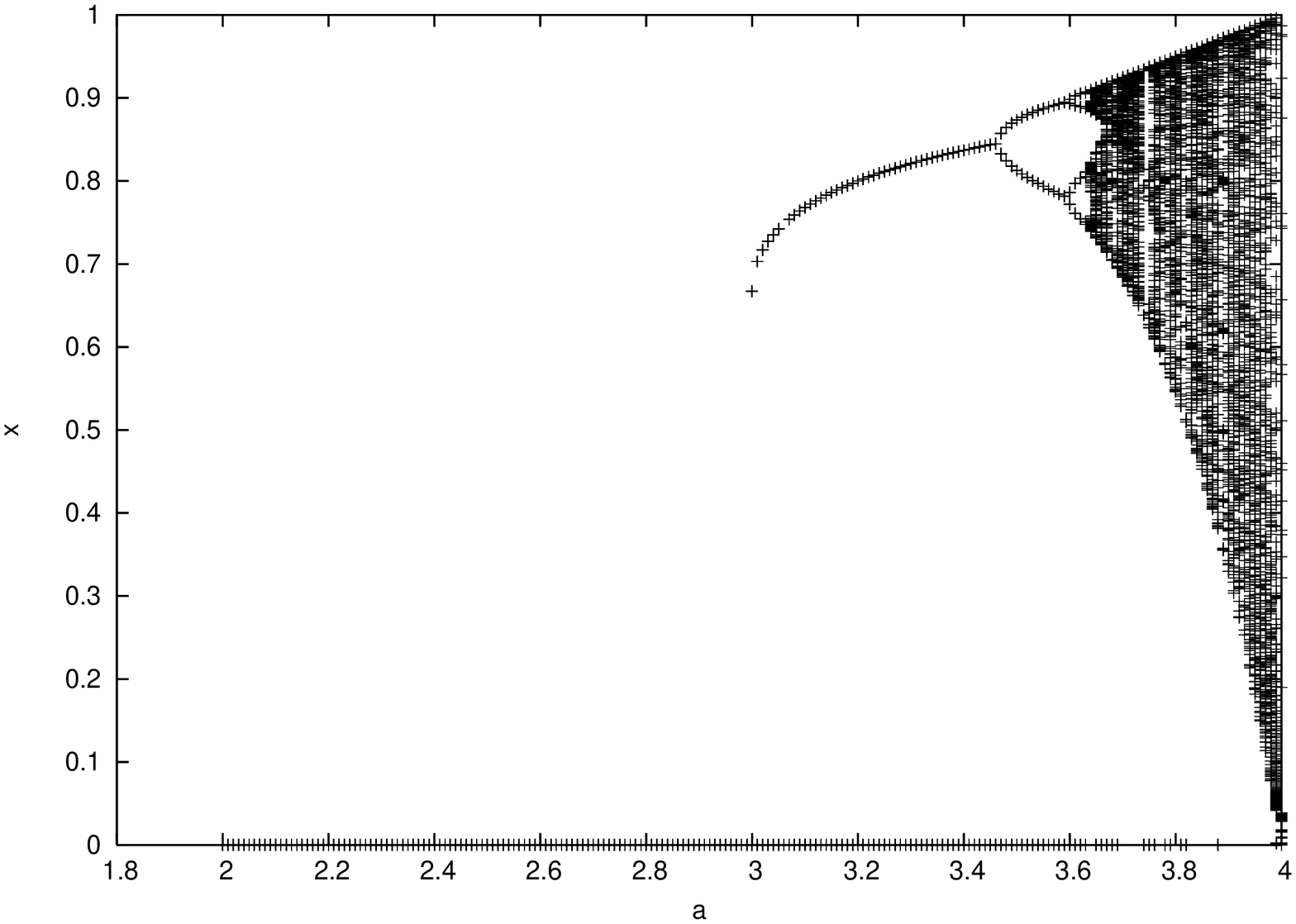,width=5truecm,angle=-90}}
\end{center}
\caption{Bifurcation diagram of the map with respect to $a$ for $\epsilon = 3$} 
\end{figure}

\begin{figure}[htb]
\label{fig15}
\begin{center}
  \mbox{\epsfig{file=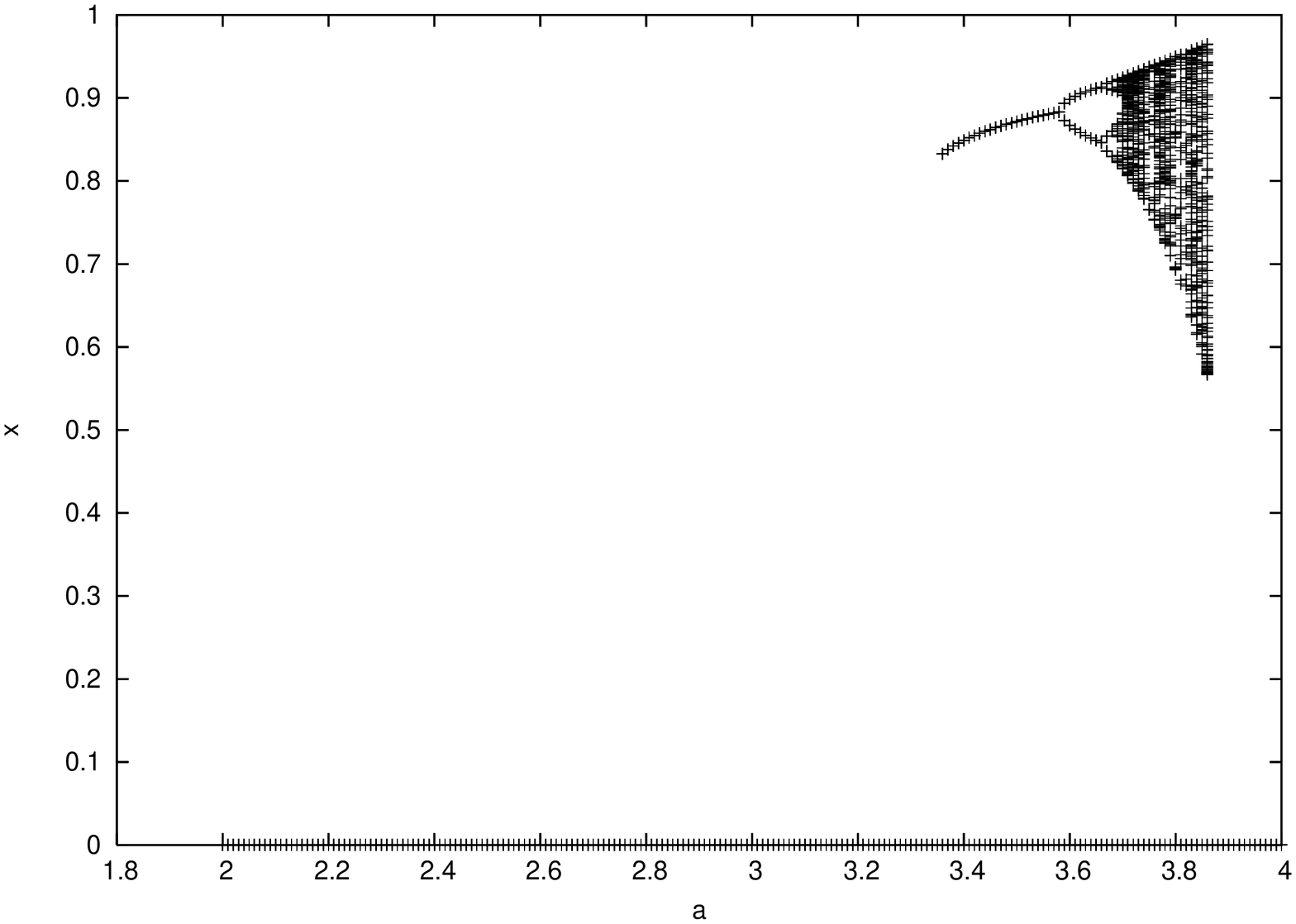,width=5truecm,angle=-90}}
\end{center}
\caption{Bifurcation diagram of the map with respect to $a$ for $\epsilon = 5$} 
\end{figure} 

Figures~16-18 display the Lyapunov exponents, obtained from trajectories 
arising from different initial conditions.  Fig.~16 displays the exponents 
with respect to $\epsilon$ for $a=3.6$, and it shows two branches of Lyapunov 
exponents arising from different initial conditions, after 
$\epsilon \sim 2.5$.  Both these co-existing exponents are below $0$, 
indicating the co-existence of two regular dynamical behaviours. This is 
bourne out in Fig.~5 -- from where it is clearly evident that after 
$\epsilon \sim 2.5$ the fixed point $x^{\star} = 0$ co-exists with a $2$-cycle.

In Fig.~17 one observes, for certain $\epsilon$, both positive and negative
Lyapunov exponents are obtained for different initial conditions. This
signals a co-existence of chaos and regular dynamics, as is clearly
bourne out by the bifurcation diagram in Fig.~6. 

From the Lyapunov exponents shown in Fig.~18 and the corresponding
bifurcation sequence in Fig.~9, it is clear that for $a = 4$, we have a global 
chaotic attractor extending over the entire interval for the usual logistic 
map ($\epsilon = 0$), but obtain a global attractor at $x^\star = 0$ for 
$\epsilon>3$. So $q$-deformation of a chaotic map can lead to stable fixed 
points.

\begin{figure}[htb]
\begin{center}
\label{fig16}
\mbox{\epsfig{file=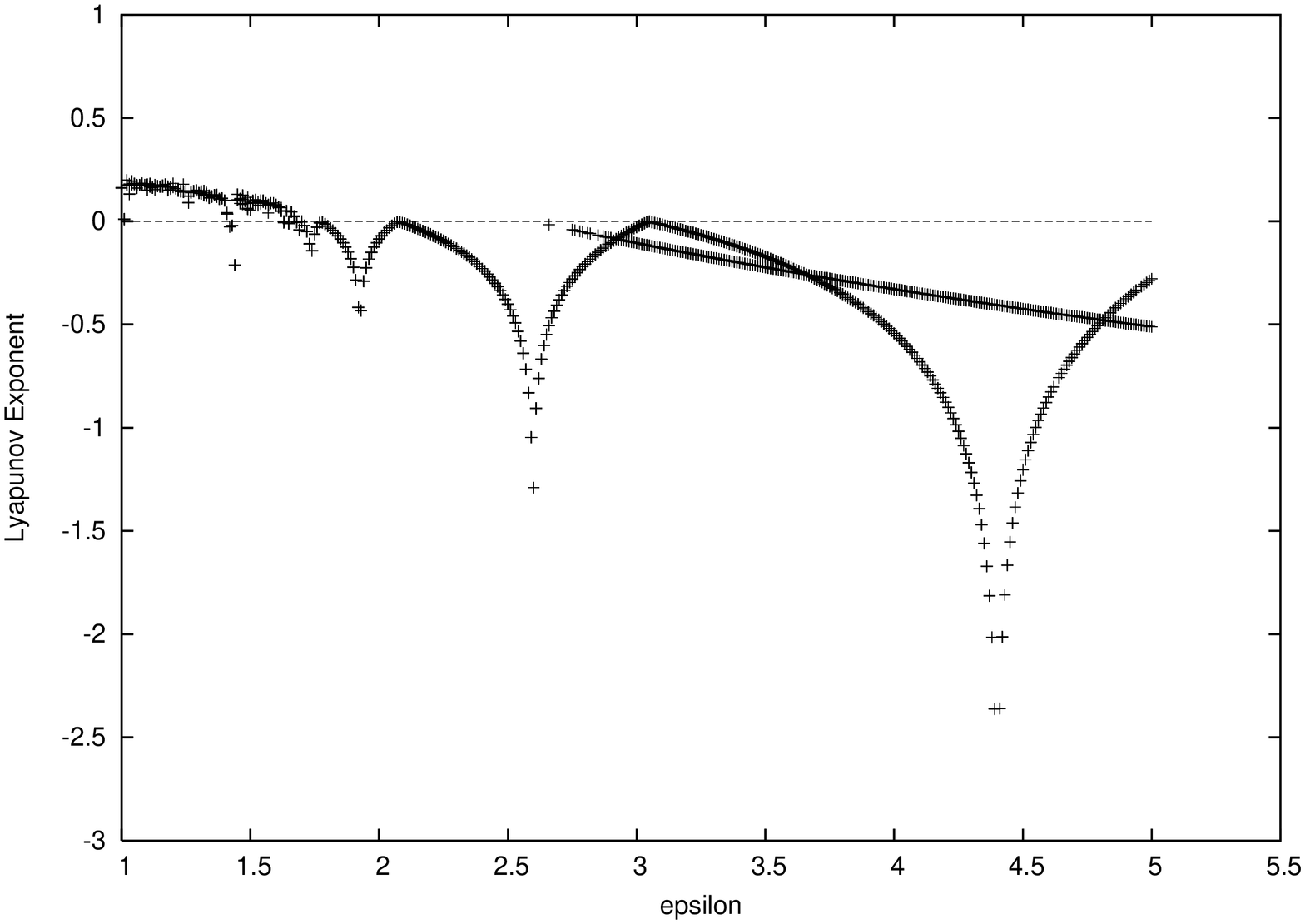, width=5truecm,angle=-90}} 
\end{center}
\caption{Lyapunov exponents with respect to $\epsilon$ for $a = 3.6$.}
\end{figure} 

\begin{figure}[htb]
\begin{center}
\label{fig17}
\mbox{\epsfig{file=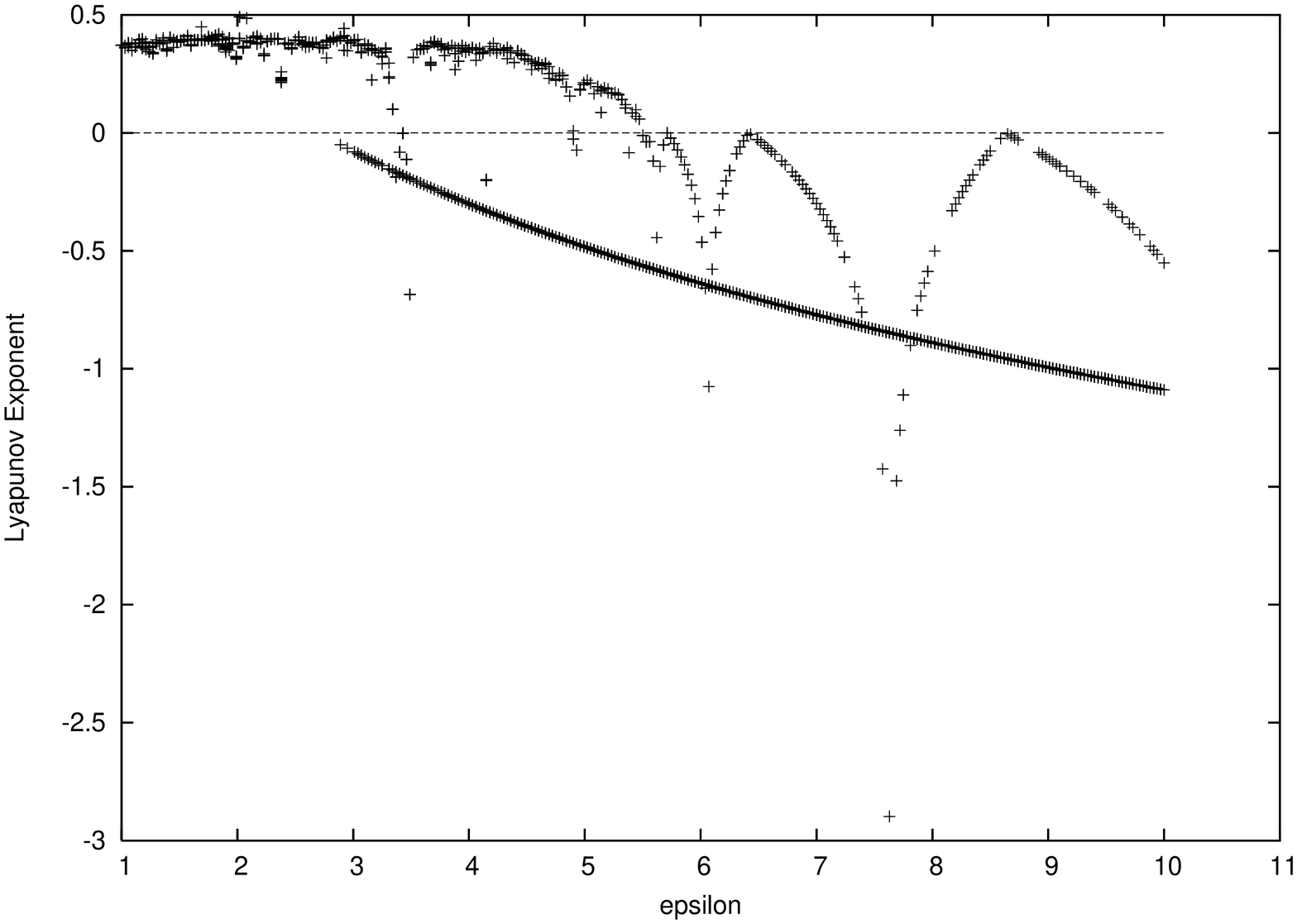, width=5truecm,angle=-90}}
\end{center}
\caption{Lyapunov exponents with respect to $\epsilon$ for $a = 3.7$.}
\end{figure} 

\begin{figure}[htb]
\begin{center}
\label{fig18}
\mbox{\epsfig{file=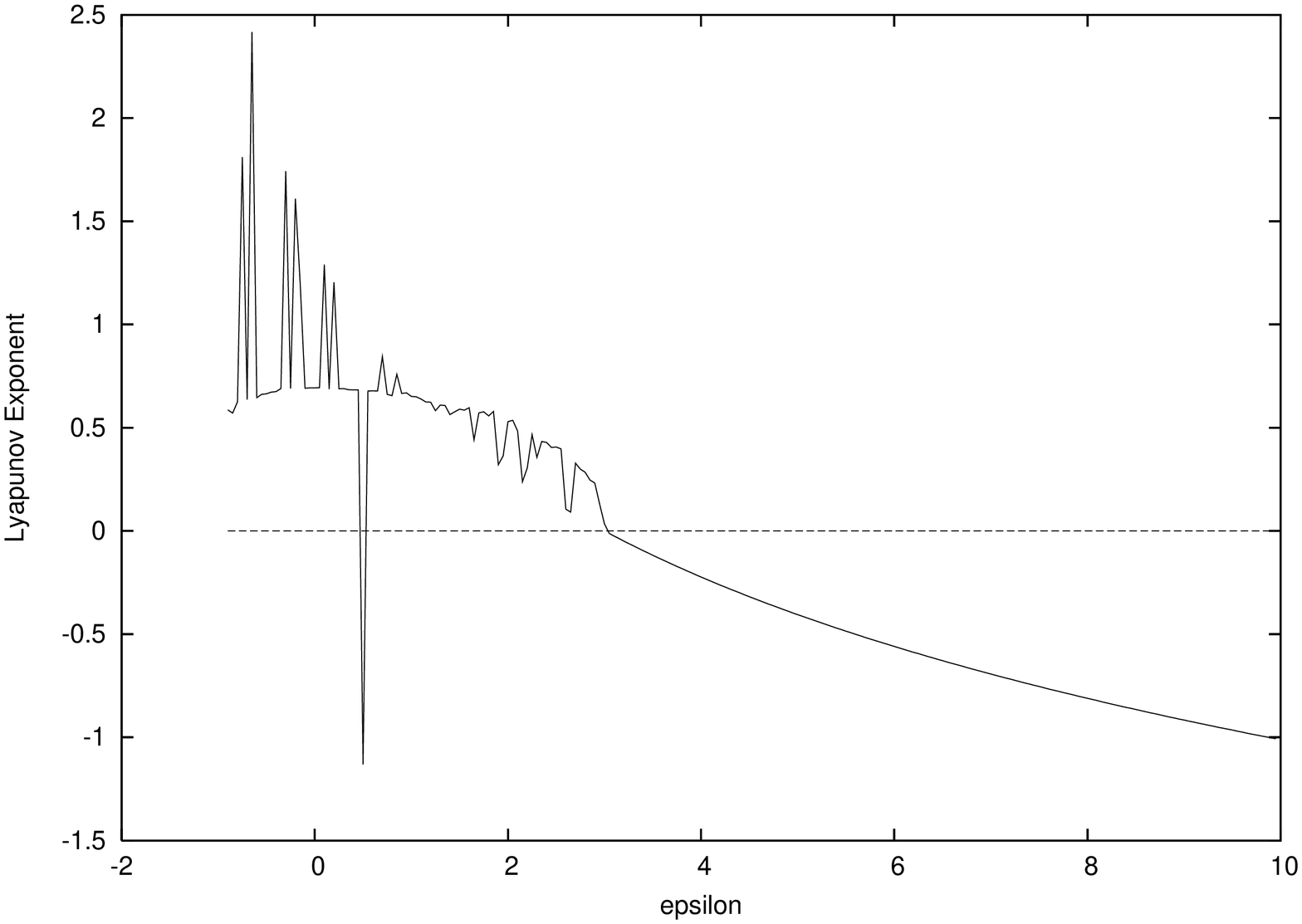, width=5truecm,angle=-90}} 
\end{center}
\caption{Lyapunov exponents with respect to $\epsilon$ for $a = 4$.}
\end{figure} 

The co-existence of attractors for the $q$-logistic map can be understood 
as follows.  Consider, for example, the case of $a = 3.7$and $\epsilon = 7$ 
where a $2$-cycle co-exists with the fixed point $x^\star = 0$.  Figure~19 
shows the form of the dynamical evolution function $F(x)$ with respect to 
$x$ for these values of $a$ and $\epsilon$.  The figure also displays the 
$F(x) = x$ line and the $F(x) = x^\star_-$ line.  The intersection of the 
former with the $F(x)$ curve yields the fixed points $x^\star = 0$, 
$x^\star_-$ and $x^\star_+$.  The intersection of the latter with the $F(x)$ 
curve yields the two pre-images for the fixed point $x^\star_-$.  One 
pre-image is simply $x^\star_-$ (i.e., it is also on the $45^o$ line) as 
it is a fixed point solution. The other pre-image is given by\,:
\begin{equation} 
F^{-1} (x^{\star}_-) 
 = \frac{(2\epsilon\epsilon^{\prime}x^{\star}_-  
      + a^{\prime}) + \sqrt{(2\epsilon\epsilon^{\prime}x^{\star}_- 
      + a^{\prime})^2 - 4 (\epsilon^2 x^{\star}_- 
      + a^{\prime})\epsilon^{\prime 2}x^{\star}_-}} 
        {2(\epsilon^2 x^{\star}_- + a^{\prime})} 
\end{equation} 
where $\epsilon^\prime = 1+\epsilon$ and $a^\prime = a\epsilon^\prime$.  
This marks the beginning of the interval $[F^{-1}(x^\star_-),1]$ which
maps on to the interval $[0,x^\star_-]$ in the consequent iteration.  These 
two intervals are dynamically connected and all points in it are attracted 
to the fixed point $x^\star = 0$. Notice however that the interval 
$[x^\star_-,F^{-1}(x^\star_-)]$ is never mapped to the other two. In fact 
it always maps on to itself.  Basically the $F(x)$ in the range 
$[x^\star_-,F^{-1}(x^\star_-)]$ will lie in the interval  
$[x^\star_-,F_{max}]$, where $F_{max}$ is the maximum of the map $F(x)$.  
Whenever $F_{max} < F^{-1}(x^\star_-)$, the middle segment will map onto 
itself.  When this happens the dynamics in this segment will be distinct 
from that in the other two, and we will have a co-existence of dynamical 
attractors.  Figure~19 also shows the numerically obtained basin of 
attraction for the fixed point $x^\star = 0$ as a black bar on the 
$x$-axis, and clearly it falls exactly in the intervals outlined above.

\begin{figure}[htb] 
\begin{center}
\label{fig19}
\mbox{\epsfig{file=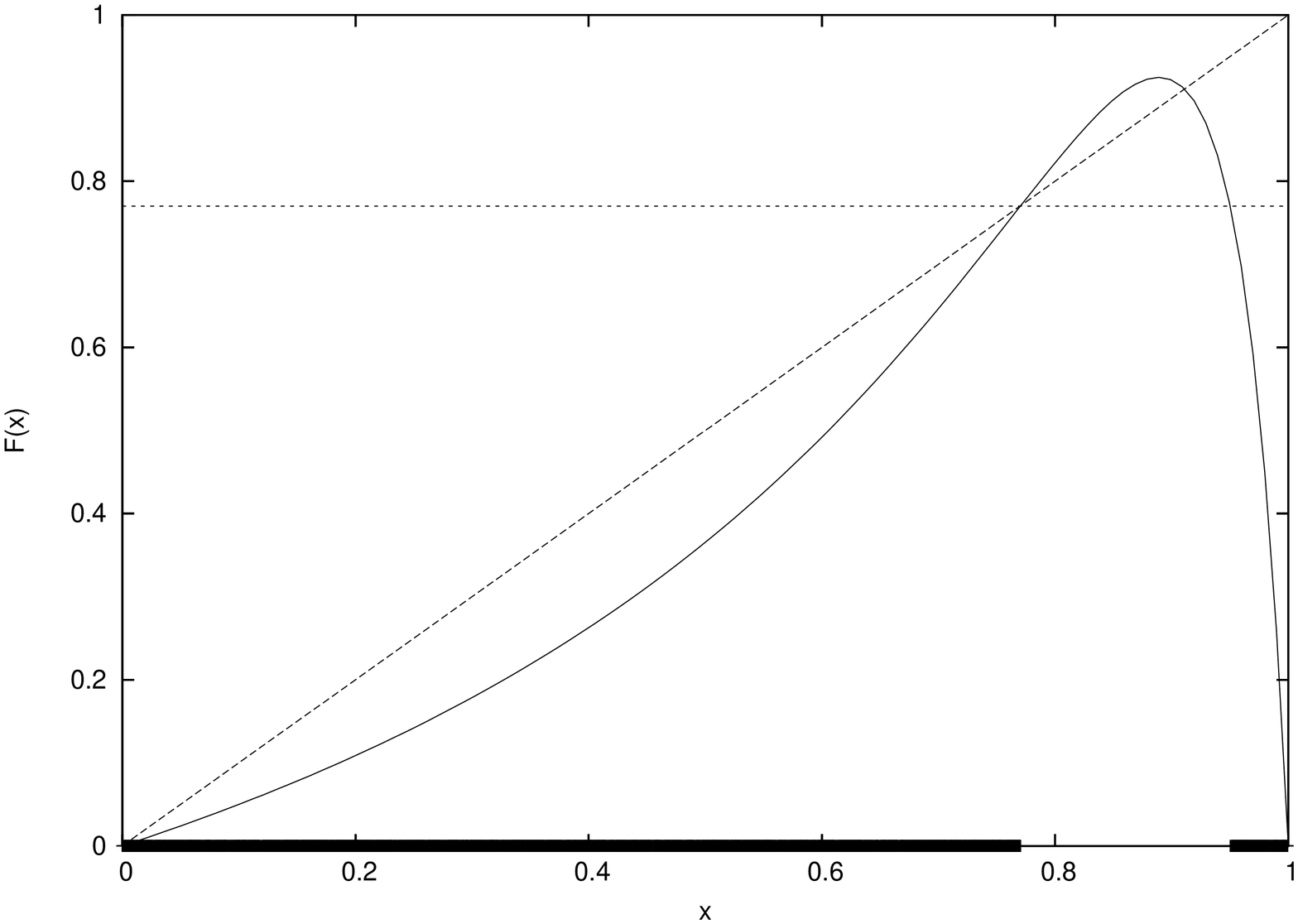,width=5truecm,angle=-90}}
\end{center}
\caption{The form of the dynamical evolution function $F(x)$ with respect to 
$x$ for $a = 3.7$ and  $\epsilon = 7$. The dashed lines are\,: the $F(x) = x$ 
line ($45^o$ line) and the $F(x) = x^\star_-$ line.  The figure also shows the 
numerically obtained basin of attraction for the fixed point $x^\star = 0$ as 
a black bar on the $x$-axis.}
\end{figure}

\section{Conclusions}  
To summarize, we have introduced a scheme of $q$-deformation of nonlinear maps, 
enthused over studies on several $q$-deformed physical systems related to 
quantum group structures and inspired by the mathematical basis of Tsallis 
statistical mechanics. We characterized the scheme in detail with reference to 
the logistic map.  The resulting family of $q$-logistic maps has been found to 
have an interestingly wide spectrum of behaviours, compared to the usual
logistic map.  In particular one observed co-existence of attractors -- a 
phenomenon rare in one dimensional maps.  

One can also study the $q$-deformations of maps following other deformation 
schemes, like based on~(\ref{eq:heinen}) or~(\ref{eq:qgn}).   We feel that 
the study of such families of deformed maps should be profitable for 
analytical modeling of several phenomena, as one parameter, namely the 
deformation parameter, can be used to fit a large range of functional 
forms, as evident from Fig.~1.  For instance, it is seen that the 
experimentally constructed one dimensional map for the Belousov-Zhabotinskii 
reaction in a stirred chemical reactor~\cite{BZ} has a striking similarity to 
the $q$-logistic map in Fig.~1 for negative epsilon value. Clearly there are 
many interesting areas of research for further exploration.  One can study 
the $q$-deformations of the various other nonlinear maps (one or higher 
dimensional) and coupled map systems.  Further, in higher dimensional cases 
and coupled map systems one can experiment with different deformation 
parameters ($\epsilon$) for different variables and for different maps coupled 
in the system.  In conclusion, $q$-deformation allows us to generate 
interesting families of maps and has potential utility in constructing 
iterated functional schemes to model low dimensional dynamical phenomena.

\end{document}